\documentclass[11pt, preprint]{aastex}
\usepackage{subfigure}
\let\tablehead\undefined
\let\tabletail\undefined
\usepackage{supertabular}
\let\affil\undefined
\usepackage{authblk}
\def\etal{~et al.}
\def\hi{H~I~}
\def\hip{H~I}
\def\logoh{$\log(O/H)~$}
\def\logohp{$\log(O/H)$}

\title{\bf Dependence of Nebular Heavy-Element Abundance on H~I Content for Spiral Galaxies}
\author[1]{Paul Robertson}
\author[1]{Gregory A. Shields}
\author[2]{Romeel Dav\'{e}}
\author[3]{Guillermo A. Blanc}
\author[1]{Audrey Wright}
\affil[1]{Department of Astronomy, The University of Texas, Austin, TX, 78712; paul@astro.as.utexas.edu}
\affil[2]{Astronomy Department, University of Arizona, Tucson, AZ 85721, USA}
\affil[3]{Carnegie Observatories, Pasadena, CA 91101-1292}
\begin{abstract}

We analyze the galactic \hi content and nebular \logoh for 60
spiral galaxies in the Moustakas et al. (2006) spectral catalog.
After correcting for the mass-metallicity relationship, we show that
the spirals in cluster environments show a positive correlation
for \logoh on DEF, the galactic \hi deficiency parameter, extending
the results of previous analyses of the Virgo and Pegasus I
clusters.  Additionally, we show for the first time that galaxies in
the field obey a similar dependence.  The observed relationship between
\hi deficiency and galactic metallicity resembles similar trends shown by
cosmological simulations of galaxy formation including inflows and outflows.  These results indicate the previously observed metallicity-DEF correlation has a more universal interpretation than simply a cluster's effects on its member galaxies.  Rather, we observe in all environments the stochastic effects of metal-poor infall as minor mergers and accretion help to build giant spirals.

\end{abstract}


\begin{document}


\section{\bf Introduction}

The relationship between galactic \hi content and other properties of giant spiral galaxies is a well-documented phenomenon.  Most notably, star formation \citep[e.g.][]{kennicutt98,rose10} and gas-phase metal abundance \citep[][among others]{skillman96,ellison09,zhang09} are known to be intimately connected to a galaxy's overall \hi mass.  Many observational studies of the \hip-metallicity relation interpret the phenomenon as a consequence of environment-driven evolution (namely ram pressure gas stripping or infall cutoff) through either cluster membership \citep{skillman96,petropoulou12} or local overdensity \citep{cooper08,ellison09}.  On the other hand, cosmological hydrodynamical simulations \citep[e.g.][]{dave11b} predict a dependence of galactic metallicity on \hi content for galaxies in the field as well.

In \citet{robertson12}, we took the approach of the \citet{skillman96} analysis of Virgo, examining a single galaxy cluster--Pegasus I--to evaluate the effect of \hi content on mean galactic metallicity for giant spirals.  Rather than bifurcate our sample into ``gas-rich" and ``gas-poor," as had been done for Virgo, we quantified \hi content using the DEF parameter described in \citet{sgh96}, and considered galactic abundances as a function of this quantity.  In the case of Pegasus I, DEF proved to be an excellent predictor of galactic \logohp.  Furthermore, the Virgo galaxies from the \citet{skillman96} study agreed nicely with the \logohp-DEF correlation despite the dramatically different properties (density, number of galaxies, velocity dispersion) of the two clusters.

The most significant limitation of the observed relation between \logoh and DEF is the small number of galaxies for which the dependence has been tested.  Between the Virgo and Pegasus I clusters, only 12 cluster galaxies were included in the \citet{robertson12} paper.  Furthermore, while we included a small number of field spirals from the \citet{zkh94} sample, the number of objects and the precision of their associated \logoh measurements made it impossible to conclude whether our observed correlation extended to galaxies in the field.  In this paper, we remedy both of these shortcomings by utilizing galaxy-integrated spectra of 60 giant spirals (35 cluster, 25 field) from the \citet{moustakas06} catalog.  Here, we show that the abundances of these galaxies confirm the dependence of galactic \logoh on DEF for cluster spirals, and that field spirals are subject to a similar relation, in agreement with cosmological hydrodynamical simulations.

\section{\bf Data}

To expand on the results of \citet{skillman96} and \citet[][hereafter Paper I]{robertson12}, we sought to obtain \hi and metallicity measures
for a large number of galaxies in a wide range of environments.
Because our \hi deficiency parameter DEF requires accurate 21 cm \hi
fluxes and morphological types, we were confined to relatively nearby
galaxies.  Also, since accurate \logoh determinations for spiral
galaxies are dependent on spectra covering the entire galactic disk,
very large surveys such as SDSS are unsuitable, as nearby spirals do
not fit within a single fiber.

We found a suitable sample of objects in the \citet{moustakas06}
catalog of long-slit galactic spectra.  The catalog contains
emission-line spectra for 417 galaxies.  While Paper I and other
similar studies determine galactic nebular metallicities by fitting
abundance gradients to spatially resolved H~II region spectra,
\citet{mk06} show that the integrated spectra from these long-slit
observations yield equivalent \logoh values.  Taking advantage of
their result, we derived galactic abundances from this catalog.
First, though, we selected the galaxies suitable for our
$\log(O/H)$/DEF analysis according to the following criteria:

I.  We selected only objects for which \hi 21cm flux measurements, optical diameters, and T-types exist
in the Third Reference Catalog of Bright Galaxies \citep[RC3][]{gdv91}.

II.  We eliminated any objects without significant detections of the
$\textrm{[O~II]} \lambda 3727$ or $\textrm{H}~\beta$ emission lines.

III.  We selected only massive spirals, with T-types between 0 and 8.
Additionally, we eliminated any objects known to be in interacting or
merging pairs because of the difficulty of assigning morphological
types to these galaxies, and because of the known metallicity dilution
effects \citep{kewley06,ellison08} for interacting pairs.  Galaxies known to be
in groups (not clusters) have also been eliminated due to their
relatively limited number.


After selecting for the above requirements, we are left with a sample
of 60 spiral galaxies.  For these objects, we first calculated the
\hi deficiency parameter DEF \citep{giovanelli85}.  We computed DEF following
\citet{sgh96}, who define the quantity as

$\textrm{DEF} = \log M_{\textrm{H~I},exp} - \log M_{\textrm{H~I}}$

\noindent where $M_{\textrm{H~I},exp}$ is an expectation value for a
galaxy's \hi mass based on its optical diameter and morphological
type.  Since DEF is an underabundance relative to the expectation,
more positive values represent lower \hi content.
  
As in Paper I, we used oxygen as a proxy for a galaxy's heavy-element
abundance, and used the strong-line $R_{23}$ calibration for the
[O~II] and [O~III] emission lines.  To facilitate direct
comparison to Paper I, we have again used the \citet{zkh94} $R_{23}$
calibration to compute $12 + \log(O/H)$.  Our error bars are obtained
from standard propagation of the uncertainties given for the
\citet{moustakas06} emission lines.  

We categorized our galaxies as cluster, group, or field members using the associations listed in HyperLeda \citep{paturel03}.  In cases where HyperLeda did not offer this information, we consulted the SIMBAD and SDSS SkyServer Object Explorer databases, and references therein.  If a galaxy was not listed as a group or cluster member in any available literature or database, we considered it a field galaxy.  

In Table \ref{data}, we list the names, DEFs, $12 + \log(O/H)$
values, and, where applicable, host clusters of the galaxies examined in this study.  For cluster members, we have also included approximate sky-projected separations $\rho_C$ from the cluster center, using the coordinates and redshifts of cluster centers from \citet{bp84}, assuming $H_0 = 72$ km s$^{-1}$ Mpc$^{-1}$.  The Table is separated into cluster and field populations, as they will be presented in the
following section.

\section{\bf Analysis}

\begin{figure}
  \begin{center}
    \includegraphics[scale=0.6]{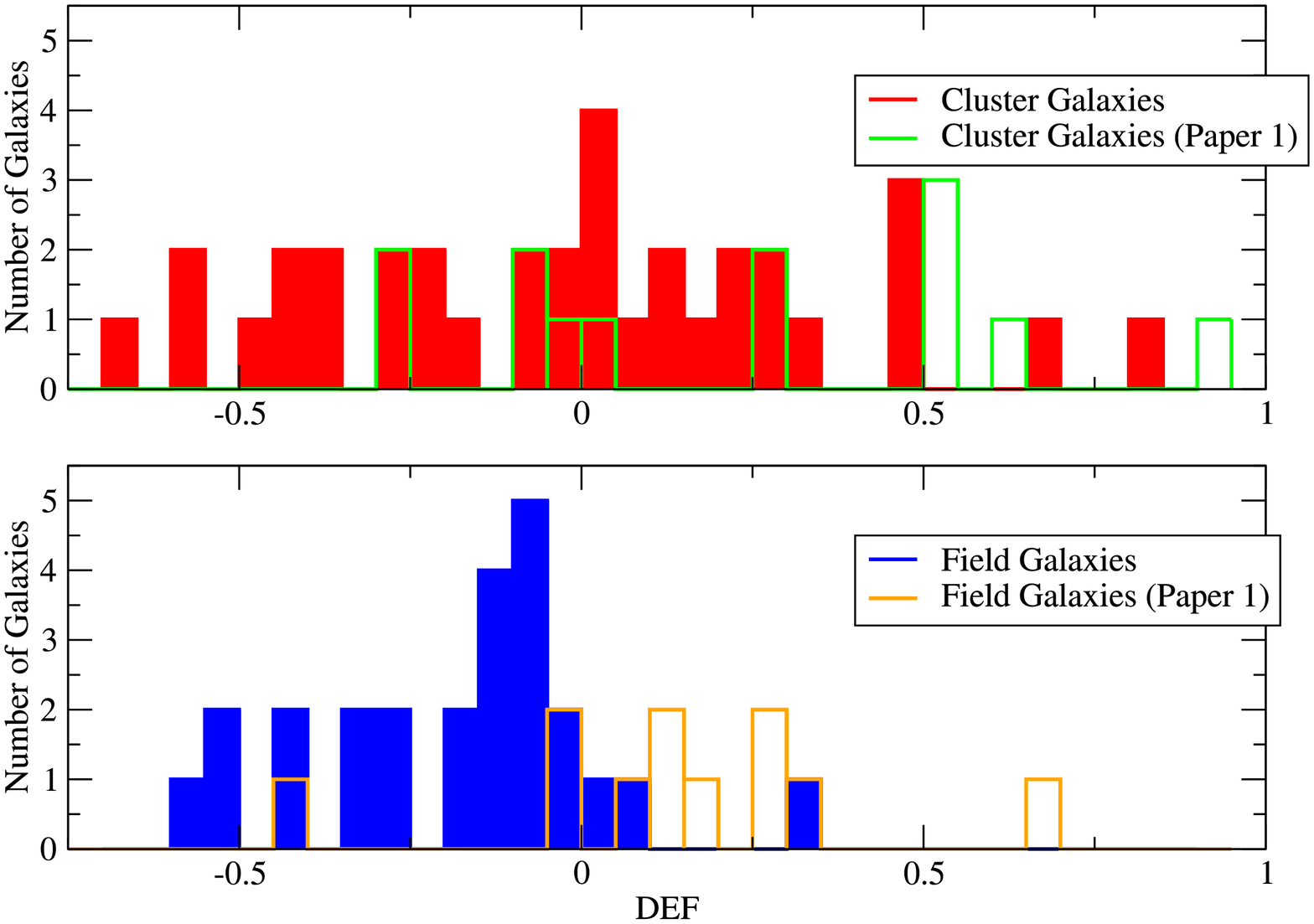}
    \caption{Distributions of the \hi deficiency parameter DEF for our
      cluster (top) and field (bottom) selections of galaxies from the \citet{moustakas06} catalog.  The open bins show the DEF distributions from Paper I.}
    \label{defhist}
    \end{center}
\end{figure}

As in Paper I, we are interested in the dependence of \logoh on DEF for the galaxies in Table \ref{data}.  In order to evaluate any functional dependence, it is important that our sample cover a satisfactory dynamical range in DEF.  In Figure \ref{defhist}, we show a histogram of DEF for the galaxies studied herein.  For comparison, we also indicate the DEFs sampled in Paper I.  We see that these objects cover a broad range of \hi deficiency, and include significantly more very high- and low-DEF galaxies than the targets of Paper I and \citet{skillman96}.  We note that, while there are members of both cluster and field samples with very low DEFs, there are considerably more cluster galaxies with positive DEF values.  This is consistent with the results of \citet{solanes01} and \citet{levy07}, among others, who show that the cluster environment drives galactic \hi depletion.

In order to properly understand the influence of \hi content on
galactic heavy-element abundance, we must first correct for the
mass-metallicity relationship \citep[MZR,][]{zkh94,tremonti04}.
To ensure easy comparison to Paper I, we have again removed the effect of the
MZR by using inclination-corrected circular velocity as a proxy for
galactic mass, and subtracting the \logoh versus $v_C$ fit derived in
Paper I:

$12 + \log$(O/H)$ = 8.57+0.356 \times V_C/(200~\mathrm{km/s})$.

We plot the residual \logoh differential for each galaxy in Figure
\ref{ovdef}.  In order to ensure that our results are not dependent on our MZR correction, we also present the same data, corrected by instead subtracting the \logoh versus $M_B$ relation from Paper I:

$12 + \log$(O/H)$ = 8.95-0.0594 \times (M_B+20)$.

Note that we show our $M_B$-corrected data as a consistency check, and base all of our formal conclusions on the $v_C$-based MZR correction.  This is because, as mentioned in Paper I and \citet{zkh94}, $v_C$ is independent of distance and unbiased by recent star formation.

Because Paper I showed a clear correlation between
oxygen content and DEF for cluster galaxies, but was unable to confirm
or reject that correlation for galaxies in the field, we examine the
cluster and field galaxies separately.

\subsection{Cluster Galaxies}

\begin{figure}
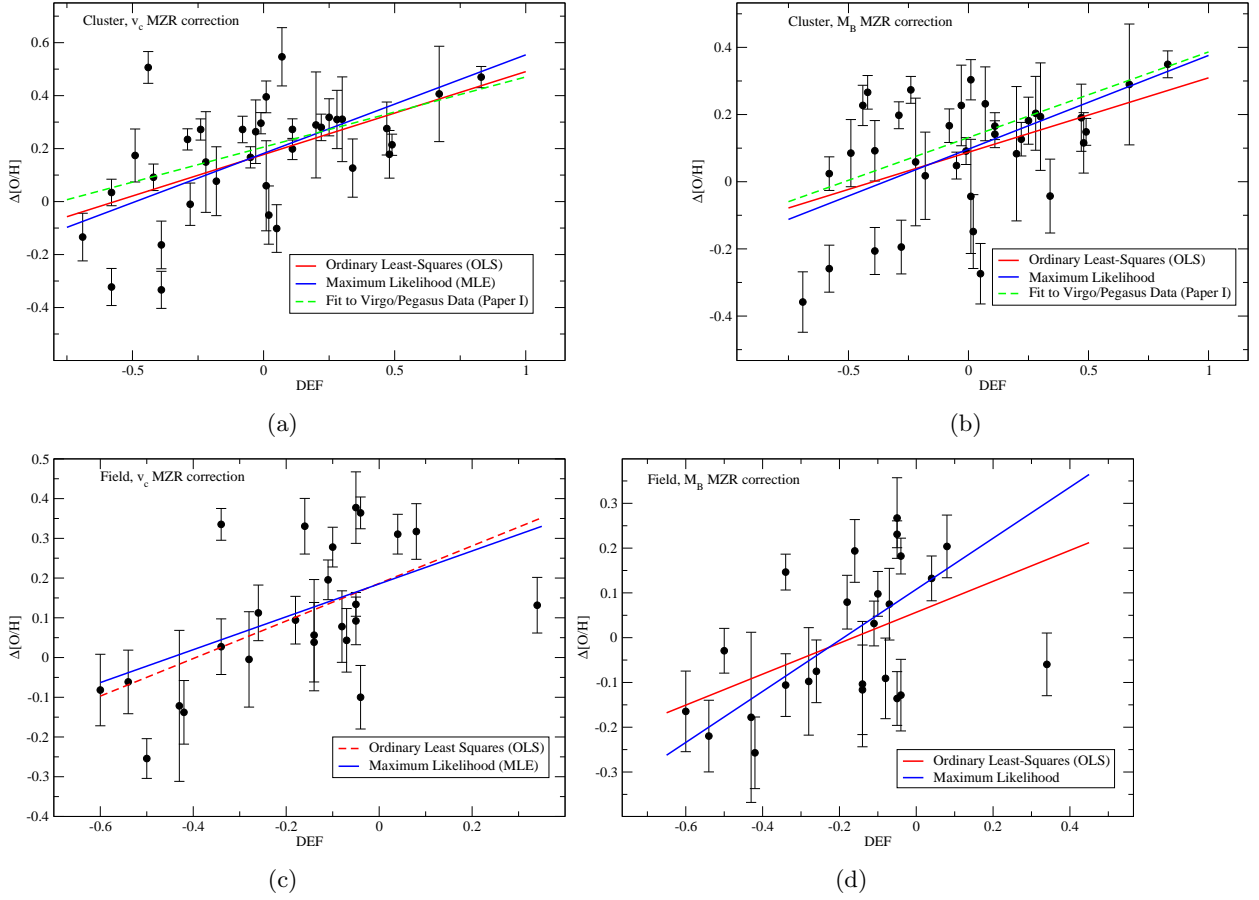

\subfigure[\label{ovdef_cluster}]{\includegraphics[width=0.45\columnwidth]{vDiffvDef_rc3_cluster.eps}}
\subfigure[\label{movdef_cluster}]{\includegraphics[width=0.45\columnwidth]{mDiffvDef_rc3_cluster.eps}}
\subfigure[\label{ovdef_field}]{\includegraphics[width=0.45\columnwidth,clip=true]{vDiffvDef_rc3_field.eps}}
\subfigure[\label{movdef_field}]{\includegraphics[width=0.45\columnwidth,clip=true]{mDiffvDef_rc3_field.eps}}
\caption{Residual \logoh after subtracting the mass-metallicity
  relationship (MZR) for our selected galaxies, plotted as a function of
  DEF.  Our sample is separated into [a,b] cluster and [c,d] field
  galaxies.  Plots on the left [a,c] have been corrected for the MZR using circular velocity, while plots on the right [b,d] use absolute blue magnitude instead (see text for details).  For each subset of galaxies, we have included linear fits
according to ordinary least squares (red) and maximum likelihood (blue).  For the cluster galaxies, we have also included our ordinary least squares fit to the Virgo/Pegasus data from Paper I (dashed green line).}
\label{ovdef}
\end{figure}

Considering first the subset of cluster galaxies (Figure
\ref{ovdef_cluster}), we see that the greatly increased number of
objects contains a considerable amount of scatter in comparison to the
Virgo/Pegasus sample of Paper I (Figure 6 in that paper).  Evaluating the
dependence of \logoh on DEF therefore requires a careful statistical
analysis.

As mentioned in the previous section, uncertainties on galactic \logoh
are obtained in a straightforward manner from the errors on the line
fluxes.  However, understanding the uncertainty on DEF is considerably
more complicated.  Because the calcluation of DEF relies on T-type and
optical diameter in addition to 21 cm flux, uncertainties in all of
those parameters contribute to the overall error budget.
Additionally, since DEF is calibrated to a finite sample of field
galaxies \citep{sgh96}, the calculation of expected \hi mass is not
exact.  Rather than assign individual errors to each object, we chose
instead to adopt a uniform error $\sigma_{\footnotesize{\textrm{DEF}}} = 0.15$ for
all galaxies, based on the recommendation of \citet{levy07}, who
estimate a ``cosmic scatter'' of 0.15 in DEF.  Our derived
dependencies on DEF will therefore have relatively conservative error
estimates, since purely statistical error calculations would result in smaller uncertainties.

We began our analysis with a standard linear regression on the cluster
subset.  Although we experimented with a number of weighting schemes, because
the uncertainties in our data only differ in the estimates of
$\log(O/H)$, which does not by itself dominate the error budget, each of
our weighted fits resulted in unreasonably small errors on the
resulting slopes and intercepts.  For this reason, all least-squares
fits presented herein are computed with equal weights for all data
points.  With an ordinary least squares (OLS) fit, our model is  

\begin{equation}
\label{ols_cluster}
\log(O/H)_{res} = 0.18_{\pm 0.03} + 0.31_{\pm 0.08} \times \textrm{DEF}
\end{equation}

\noindent where $\log(O/H)_{res}$ is the measured abundance after
subtracting our MZR fit.  

Because our fitted slope is only $\sim 3\sigma$ away from zero, we also performed a Pearson correlation test on the data in our cluster sample to confirm the statistical significance of the relation between \logoh and DEF.  The correlation coefficient for the cluster galaxies is 0.55.  For a sample size of 35 galaxies, this coefficient indicates the probability of no correlation is just 0.0006.  We see, then, that there is a significant correlation between \logoh and DEF, and the slope of the relation is consistent with that found in Paper I.





To obtain a better estimate of the actual functional relationship
between \logoh and DEF, we have performed a more statistically
rigorous linear fit to the data using the maximum likelihood (MLE) method outlined by \citet{kelly07}.  The code works by creating a likelihood function for the true distribution of regression parameters, based on the observed data and errors.  The regression coefficients and errors are estimated by performing Bayesian inference using 10,000 MCMC samples of the parameter space, where each chain performs a random walk through the parameter space (using a Gibbs sampler), eventually converging on the posterior distribution.  The values of the slope and intercept to which each chain converges represents a single random draw from the posterior distribution.  Fitting a gaussian to the resulting distribution of slopes, and extracting the mean and FWHM, the
resulting ``mean fit'' to the cluster subset becomes

\begin{equation}
\label{mle_cluster}
\log(O/H)_{res} = 0.18_{\pm 0.15} + 0.37_{\pm 0.15} \times \textrm{DEF}
\end{equation}

We have included both of the fits above in Figure \ref{ovdef_cluster}.  Since the MLE routine allowed us to include uncertainties on both DEF and \logohp, we adopt the MLE fit as our final model.  However, it is worth pointing out the agreement between the OLS and MLE fits for the cluster subset, suggesting OLS is actually adequate in this case.

When considering our sample using the $M_B$-based MZR, we find results consistent with our primary MZR correction.  We obtain slopes of $0.31 \pm 0.08$ (OLS) and $0.28 \pm 0.13$ (MLE), which agree with the fits above.

In addition to being internally consistent, our fits to the \logoh versus DEF relation, also agree with our fits to the Virgo and Pegasus spirals derived in Paper I.  For comparison, we have included these fits in Figures \ref{ovdef_cluster} and \ref{movdef_cluster}.

\subsection{Field Galaxies}

Having recovered the \logoh versus DEF relationship discovered in Paper I, we revisited the question of whether the same dependence exists for galaxies in the field.  Our field galaxy sample is plotted in Figure \ref{ovdef_field}.  Again, there is plenty of scatter, but a positive trend is visible.  We again performed a Pearson correlation test to the field sample, acquiring a correlation coefficient of 0.58.  For 25 galaxies, our correlation coefficient gives the probability of no correlation at $P = 0.0024$.  Our OLS model for the field subset gives

\begin{equation}
\label{ols_field}
\log(O/H)_{res} = 0.19_{\pm 0.04} + 0.47_{\pm 0.14} \times \textrm{DEF}
\end{equation}




To properly quantify the relationship using our errors on DEF, we again calculate the MLE fit, giving a final model

\begin{equation}
\label{mle_field}
\log(O/H)_{res} = 0.19_{\pm 0.09} + 0.41_{\pm 0.14} \times \textrm{DEF}
\end{equation}


We note that the OLS and MLE fits for field galaxies all agree to within $1\sigma$ regardless of which MZR correction we use.  However, it is worth noting that the MLE fit to the $M_B$-corrected data (Figure \ref{movdef_field} results in a very steep slope of $0.57 \pm 0.4$.  We find that when we exclude NGC 4605, which is very metal-poor for its DEF value (0.34), the OLS fit also displays a much higher slope.  Since our $v_C$ MZR correction places NGC 4605 in better agreement with the observed trend, we do not exclude it as an outlier.

Since the \citet{moustakas06} selection of spiral galaxies is not a volume-limited sample, it is prudent to consider whether our observed trends in \logoh versus DEF could be produced by an observational bias.  The RC3 catalog is essentially complete for galaxies with optical diameters greater than 1 arcminute and total $B$ magnitudes brigher than 15.5.  Although the surface brightness cutoff may lead to the omission of some edge-on spirals, such a bias should not have a significant influence on our result, as the uncertainty in morphological type and increase in interstellar reddening complicate the determinations of DEF and \logohp, respectively.  In Paper I, we intentionally avoided edge-on spirals for this reason.

As for the \citet{moustakas06} selection, which is described in \citet{mk06} while it is neither blind nor complete, the galaxies included cover a wide range in $M_B$, $B-V$, and morphological type.  We therefore expect a representative sampling of different galaxy masses, star formation histories, and dust content.

In order for a bias to create such an effect, we would somehow have to systematically exclude \hip-rich galaxies with high metallicity and/or \hip-poor galaxies with low metallicities.  We believe both possibilities are very unlikely.  Galaxies with low DEF (high \hi content) should produce strong 21 cm radiation, and will also likely have relatively high specific star formation rates \citep[e.g.][]{rose10}, leading to strong H~$\beta$ lines.  Therefore, low-DEF galaxies should not be excluded from our selection at any metallicity, according to our criteria listed in the previous section.  As for high DEF/low metallicity spirals, their low metal content should lead to strong nebular emission lines via higher temperatures, ensuring their inclusion from the \citet{moustakas06} catalog.  Furthermore, even our highest-DEF galaxies represent 21 cm detections well above the $100\sigma$ level, so we are not excluding any high-DEF galaxies due to nondetections of \hi emission.  We are therefore confident that our result is not due to an observational bias, despite the fact that our sample was not specifically chosen to be completely unbiased.

\section{\bf Discussion}

\subsection{Comparison to Previous Observations}

The slope of the DEF-\logoh relation for cluster galaxies ($0.37 \pm 0.15$) remains in good agreement with the slope derived in Paper I ($0.26 \pm 0.1$) upon increasing the number of galaxies in our sample by a factor of 3.  While the scatter around the fit in Figure \ref{ovdef_cluster} is higher than seen in the Virgo and Pegasus samples, it is important to consider the differences between
the galaxies examined in the two studies.  The Virgo and Pegasus
galaxies selected by \citet{skillman96} and Paper I were chosen for
their abundance of bright individual H~II regions, and were also all
nearly face-on.  Furthermore, the Virgo/Pegasus galaxies were all very
similar in mass and luminosity.  In this larger sample, there is certainly scatter introduced by inclination effects, ambiguous morphological types, and
imperfect mass correction which was largely inconsequential in the
smaller, more homogeneous earlier data sets.  In this sense, the Virgo/Pegasus galaxies can be interpreted as the ``ideal case,'' and our analysis of the \citet{moustakas06} sample extends the preliminary results to a much broader group of objects.

Considering most studies of the interrelation between galactic gas content and metallicity \citep[e.g.][]{skillman96,ellison09,petropoulou12} have examined \hi deficiency in the context of cluster environment or local galactic density, it is somewhat surprising to find that the observed metallicity dependence extends to field galaxies.  In fact, our measured slope for the field subset is actually higher than for the cluster galaxies, although it is doubtful that difference is significant.  While the distributions of the slopes in our Monte Carlo resamplings are different, given the uncertainties on our fits, we are not confident that the difference in the observed slopes is significant over the range of DEF explored here.  We therefore conclude that, within the uncertainties, the \logoh versus DEF relation applies generally to any  non-interacting massive spiral galaxy in a similar way, regardless of environment.

\subsection{Comparison to Hydrodynamic Models}

Modern cosmological hydrodynamic simulations can predict the neutral hydrogen and oxygen content for representative samples of galaxies.  Here we compare our DEF-log$(O/H)_{res}$ results to the simulations of \citet{dave11b}.  Since these simulations have a box length of $48~h^{-1}$ Mpc on a side that does not contain any cluster-sized objects, their simulated sample is most appropriately compared to our field sample.  Similar to \citet{dave11b}, we have excluded any galaxies with stellar masses lower than $M_* = 2 \times 10^9 M_{\odot}$.  While the sample does not include morphological data, the masses, star formation rates, and gas fractions of our simulated galaxies are a good match to the selection of \citet{moustakas06}, who note in \citet{mk06} that their observed galaxies are largely late-type (Sbc and later) spirals.

\begin{figure}
\begin{center}
\includegraphics[scale=0.5]{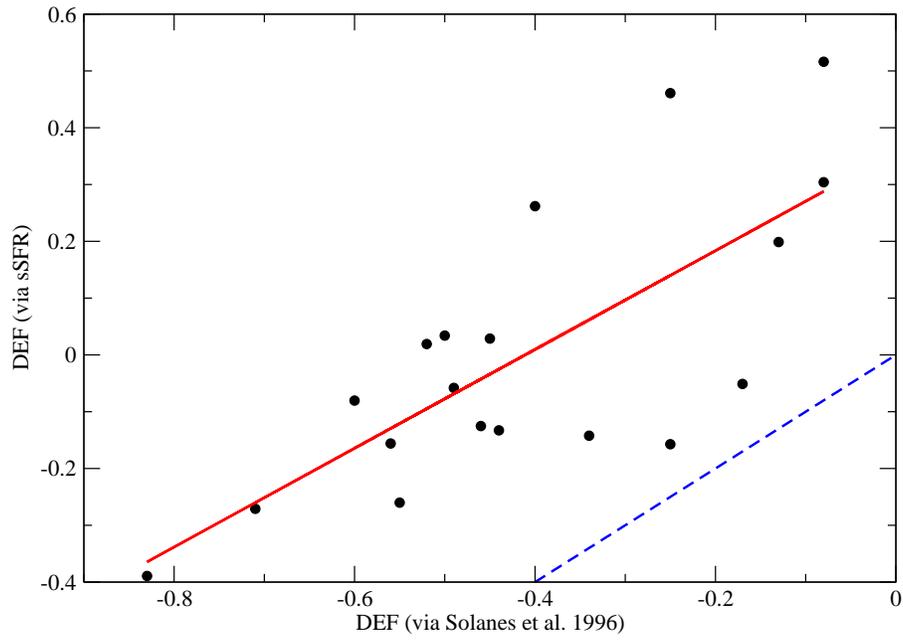}
\caption{We compare DEF as measured by the \citep{sgh96} method and by using our estimate relative to a given sSFR.  The red line gives the best fit to the data, while the dotted blue line indicates the line $y = x$.}
\label{defcomp}
\end{center}
\end{figure}

In these models, we compute the deviations in metal and \hi content at a given {\it specific star formation rate} (sSFR$\equiv {\rm SFR}/M_*$).  This is different than our treatment of the observations, where DEF is defined based on the expected HI content of galaxies with similar morphology and size.  Unfortunately, these simulations lack the resolution to predict these parameters, and hence we must choose a proxy from among the available model-predicted quantities.
We choose sSFR because \citet{rose10} showed that DEF is most tightly correlated with sSFR, as opposed to SFR or $M_*$ alone.  To verify this approach is qualitatively valid, we have examined 19 spiral galaxies with measured sSFRs from \citet{howell10}.  Taking T-types, 21cm fluxes, and optical diameters from HyperLeda, we computed DEF via the \citet{sgh96} formulae.  We then estimated \hi deficiency by correcting for a trend in $M_{\textrm{H~I}}$ versus sSFR, and adopting DEF as the vertical offset from this trend.  We show a comparison between the two estimates of DEF in Figure \ref{defcomp}.  Performing a linear fit to the data, we find a slope of $0.87 \pm 0.19$, but also a vertical offset of $0.36 \pm 0.09$, indicating a systematic difference between the two calculations.  Therefore, while we are confident that the two estimates of DEF reflect qualitatively similar trends, the offset prevents us from directly equating the simulations and our data.

\begin{figure}
\begin{center}
\includegraphics[trim = 0cm 0cm 0cm 0cm]{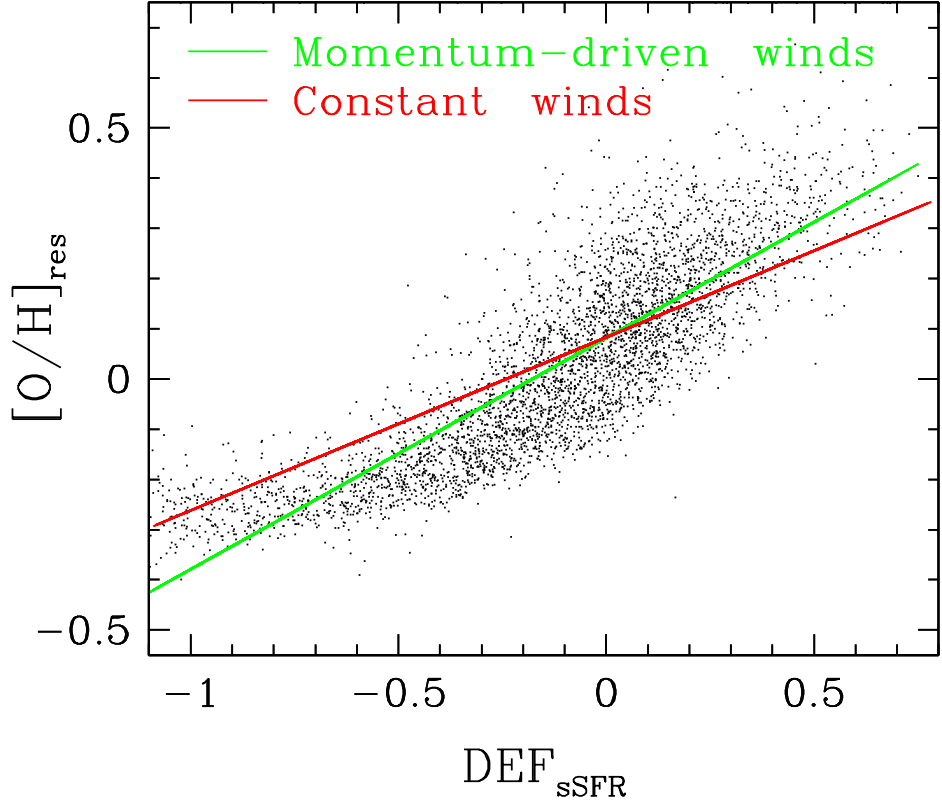}
\caption{Residual [O/H] as a function of DEF for galaxies from our hydrodynamical simulation.  The green line gives our best fit to the relation, while the red line represents the best fit to the same galaxies with a constant wind model (see text).  We note that DEF in this figure is computed relative to a ``normal" \hi content at fixed sSFR to account for a lack of morphological information in our simulations.}
\label{HIsim}
\end{center}
\end{figure}

Figure~\ref{HIsim} shows the correlation between DEF and log$(O/H)_{res}$ from the momentum-driven wind scaling simulation of \citet{dave11b}, defined relative to the mean at a given sSFR. The galaxy metallicities are computed as described in \citet{dave11b}, while the \hi mass accounts for self-shielding and conversion to molecular hydrogen as described in \citet{dave13}, broadly following \citet{popping09} and \citet{duffy12}. The green line shows the best-fit power law to these points, which follows the relation DEF$=0.07+0.43\times [O/H]_{res}$.  Also shown in the Figure is the best linear fit to the constant wind model \citep[red line, see][for details on the constant wind model]{dave11b}.

The predicted slope is close to that observed (0.41), and the results display a similar amount of scatter around the fit. The predicted amplitude is slightly low, but likely within uncertainties given the different way in which DEF is computed between the models and the data.  We note that had we chosen stellar mass rather than sSFR about which to measure our deviations, the predicted slope would be shallower, namely 0.27, but still within $1\sigma$ of that observed.  Hence \emph{the trend in DEF vs. $[O/H]_{res}$ appears to be a relatively robust prediction of hierarchical galaxy formation simulations, regardless of the details of feedback}.

Why do these hierarchical models predict such a trend?  The physical origin can be explained by appealing to the equilibrium model of galaxy evolution~\citep{dave12}.  In this scenario, galaxies live in a slowly-evolving balance between accretion, outflows, and star formation.  This results in preferred equilibrium relations for the main physical properties of galaxies, such as tight relations between stellar mass, star formation rate~\citep{dave08}, metallicity~\citep{finlator08}, and \hi content \citep{popping09}.

Galaxies are perturbed off these equilibrium relations owing to the stochasticity in accretion (e.g. mergers), which governs the scatter around these relations~\citep{finlator08}.  Consider a galaxy undergoing a merger with a smaller system.  Its metallicity will go down because the smaller system will tend to have lower metallicity. However, its \hi content will rise since smaller systems tend to be more \hip-rich.  Hence deviations towards low metallicity will be correlated with deviations towards high \hi content.  The converse can also happen, where a galaxy experiences a lull in accretion (or a dimunition owing to it becoming a satellite in a larger halo), in which case it will consume its available gas reservoir, increase its metallicity, and lower its \hi content.  It is straightforward to see that such perturbations will produce a trend in DEF vs. $[O/H]_{res}$ that is qualitatively as observed.  Furthermore, the fact that an upward trend exists regardless of wind model indicates the behavior does not arise as the result of an outflow effect, but rather it appears because of inflow stochasticity, which is independent of outflows.


We attribute the slope of the DEF vs. $[O/H]_{res}$ relation in the simulations primarily to three physical phenomena: First, it reflects the characteristic spectrum of mergers and smooth accretion that drive perturbations off the equilibrium relations.  Second, it reflects the tendency of minor mergers to enrich spirals with metal-poor gas, decreasing the global nebular metallicity.  Finally, it reflects the trend of \hi richness vs. sSFR, which analogously sets the typical deviation in \hi content when a giant spiral merges with a gas-rich dwarf.

The agreement between the models and the data suggests that the simulations are properly capturing these phenomena.  As shown in \citet{dave11b}, this model produces roughly the correct mass-metallicity relation.  In \citet{dave13} we show that it also broadly matches the observed mass-\hi richness relation.  The spectrum of mergers is set by the underlying cosmology, which is assumed to be WMAP7-concordant.  Given that all the individual pieces in the model agree with data, it is perhaps not surprising that the DEF vs. $[O/H]_{res}$ is also reproduced.  Also, since the constant and momentum-driven wind models yield qualitatively similar trends (Figure \ref{HIsim}) for metallicity and \hi richness, it is also not surprising that our simulation results are not strongly sensitive to the assumed feedback model.

We see that this scenario for what sets the \hi deficiency in galaxies offers a mechanism to produce the \logohp-DEF relation separate from traditional scenarios that have posited that it arises from environmentally-driven processes such as ram pressure stripping \citep[e.g.][]{gunn72}.  Naively, such scenarios would predict that the trends would be stronger in clusters, but our observations suggest that the trend of DEF vs. $[O/H]_{res}$ is similar in the field.  In the simulations, environment does not play a large role~\citep[except for satellite galaxies;][]{dave11a}.  Instead, DEF is simply set by the stochastic nature of hierarchical accretion, and galaxies' response to such stochasticity generically yields the observed trend in DEF vs. $[O/H]_{res}$.  We caution that these simulations only produce field galaxies, so environmental processes may still play a major role in extreme environments such as clusters.  But the success of these models suggests that at least for typical field galaxies, it is not necessary to appeal to environmental processes in order to understand the behavior of $[O/H]_{res}$ vs. DEF.  Evidently, changes to a galaxy's nebular metallicity caused by varying \hi content are to some degree insensitive to the specific physical processes (i.e. infall, minor mergers, ram-pressure stripping) responsible for regulating \hi richness.

\section{\bf Conclusion}

Using the spectral library of \citet{moustakas06}, we have conducted an expanded investigation into the influence of \hi abundance on galactic nebular metallicity analogous to the analysis of \citet{robertson12} for the Pegasus cluster.  We have compared these results to predictions based on cosmological hydrodynamical simulations.  Our conclusions can be summarized in three main results:

\emph{1.}  For galaxies in clusters, we recover the previously observed trend of increasing \logoh with decreasing \hi content.

\emph{2.}  For galaxies in the field, \logoh is similarly dependent on \hi deficiency.

\emph{3.}  Our hydrodynamical simulations for field galaxies predict a metallicity-DEF correlation similar to that observed.  We interpret this result as the product of a galaxy's natural ``excursions" between \hip-rich/metal-poor and \hip-poor/metal-rich in response to stochastic fluctuations in the inflow rate.  These departures from equilibrium with respect to the mass-metallicity relation can occur in any environment, and do not require cluster membership or enhanced local galaxy density.

\begin{acknowledgements}
We thank the anonymous referee for valuable comments.  P.~R. is supported by a University of Texas Continuing Fellowship.  G.S. gratefully acknowledges the support of the Jane and Roland Blumberg Centennial Professorship in Astronomy.  This research has made use of the SIMBAD database, operated at CDS, Strasbourg, France.  We acknowledge the usage of the HyperLeda database (http://leda.univ-lyon1.fr).
\end{acknowledgements}

\clearpage

\clearpage

\begin{center}

\tablecaption{Galaxy data from \citet{moustakas06}.  DEF has been
  computed according to \citet{sgh96}, and $12 + \log(O/H)$ is
  calibrated using the method outlined in \citet{zkh94}.  $v_C$ values are
  taken from the HyperLeda database, and are corrected for inclination.  Where appropriate, UGCl cluster listings have been replaced with their more familiar names according to \citet{bp84}.}
\label{data}	

\tablefirsthead{\hline
Galaxy Name &  DEF  &  $12 + \log(O/H)$ &  $v_C$ & Cluster & $\rho_C$ (kpc) \\
\hline}
		
\tablehead{\hline
\emph{Table \ref{data} cont'd.} & & & & &\\ \hline
Galaxy Name &  DEF  &  $12 + \log(O/H)$ &  $v_C$ & Cluster & $\rho_C$ (kpc) \\
\hline}
		
\tabletail{\hline}
\footnotesize
\begin{supertabular}{| l l l l l l |}
\multicolumn{6}{| c |}{\emph{Cluster Galaxies}} \\
\hline & & & & & \\
NGC 0660 & 	 $ -0.22 $ & 	 $ 8.97 \pm 0.19 $ & 	 140.82 & UGCl 029 & 1650 	 \\
UGC 01281 & 	 $ 0.30 $ & 	 $ 8.97 \pm 0.16 $ & 	 50.11 & UGCl 032 & 2340	 \\
UGC 01385 & 	 $ -0.29 $ & 	 $ 9.21 \pm 0.04 $ & 	 227.69 & Abell 262 & 513	 \\
NGC 0784 & 	 $ 0.34 $ & 	 $ 8.77 \pm 0.11 $ & 	 41.31 & UGCl 032 & 3700 	 \\
NGC 0877 & 	 $ -0.58 $ & 	 $ 9.09 \pm 0.05 $ & 	 272.82 & UGCl 035 & 2090	 \\
NGC 0976 & 	 $ -0.39 $ & 	 $ 9.12 \pm 0.09 $ & 	 400.9 & UGCl 038 & 6840	 \\
NGC 0972 & 	 $ 0.22 $ & 	 $ 9.11 \pm 0.05 $ & 	 145.99 & UGCl 038 & 3640 	 \\
NGC 1003 & 	 $ -0.28 $ & 	 $ 8.73 \pm 0.08 $ & 	 95.49 & Perseus & 6580	 \\
NGC 1058 & 	 $ -0.44 $ & 	 $ 9.10 \pm 0.06 $ & 	 13.27 & Perseus & 6750 	 \\
NGC 1087 & 	 $ -0.01 $ & 	 $ 9.08 \pm 0.04 $ & 	 120.27 & UGCl 043 & 937 	 \\
NGC 1345 & 	 $ -0.18 $ & 	 $ 8.82 \pm 0.13 $ & 	 97.19 & Eridanus & 369	 \\
NGC 2893 & 	 $ 0.01 $ & 	 $ 9.16 \pm 0.06 $ & 	 109.36 & UGCl 148 & 505	 \\
NGC 3079 & 	 $ 0.02 $ & 	 $ 8.89 \pm 0.11 $ & 	 208.39 & UGCl 163 & 22100	 \\
NGC 3310 & 	 $ -0.39 $ & 	 $ 8.75 \pm 0.07 $ & 	 288.38 & UGCl 163 & 9170	 \\
NGC 3353 & 	 $ 0.05 $ & 	 $ 8.57 \pm 0.09 $ & 	 57.16 & UGCl 189 & 167	 \\
NGC 3504 & 	 $ 0.49 $ & 	 $ 9.13 \pm 0.04 $ & 	 194.09 & Abell 1185 & 1590	 \\
UGC 06665 & 	 $ -0.69 $ & 	 $ 8.64 \pm 0.09 $ & 	 114.58 & UGCl 231 & 3660	 \\
NGC 3913 & 	 $ 0.20 $ & 	 $ 8.92 \pm 0.20 $ & 	 34.07 & UGCl 229 & 3260 	 \\
NGC 3953 & 	 $ 0.47 $ & 	 $ 9.23 \pm 0.10 $ & 	 215.86 & UGCl 229 & 9490	 \\
NGC 3972 & 	 $ 0.67 $ & 	 $ 9.18 \pm 0.18 $ & 	 114.36 & UGCl 229 & 3830	 \\
NGC 3982 & 	 $ 0.11 $ & 	 $ 9.11 \pm 0.04 $ & 	 191.83 & UGCl 229 & 4290	 \\
NGC 4062 & 	 $ 0.28 $ & 	 $ 9.13 \pm 0.11 $ & 	 140.47 & UGCl 263 & 8531	 \\
NGC 4085 & 	 $ 0.11 $ & 	 $ 9.07 \pm 0.04 $ & 	 127.84 & UGCl 229 & 14600	 \\
NGC 4088 & 	 $ -0.08 $ & 	 $ 9.14 \pm 0.05 $ & 	 167.29 & UGCl 229 & 14200	 \\
NGC 4102 & 	 $ 0.48 $ & 	 $ 9.03 \pm 0.09 $ & 	 158.14 & UGCl 229 & 10100	 \\
NGC 4136 & 	 $ 0.01 $ & 	 $ 8.81 \pm 0.17 $ & 	 101.3 & UGCl 263 & 3880	 \\
NGC 4157 & 	 $ -0.03 $ & 	 $ 9.17 \pm 0.12 $ & 	 188.89 & UGCl 229 & 15000	 \\
NGC 4288 & 	 $ -0.45 $ & 	 $ 8.76 \pm 0.10 $ & 	 114.37 & UGCl 265 & 388 \\
NGC 4389 & 	 $ 0.83 $ & 	 $ 9.21 \pm 0.04 $ & 	 95.47 & UGCl 265 & 47.1	 \\
NGC 4414 & 	 $ -0.24 $ & 	 $ 9.23 \pm 0.04 $ & 	 217.83 & UGCl 267 & 19.2	 \\
NGC 5014 & 	 $ 0.25 $ & 	 $ 9.04 \pm 0.07 $ & 	 85.29 & UGCl 281 & 2530	 \\
NGC 6052 & 	 $ -0.58 $ & 	 $ 8.77 \pm 0.07 $ & 	 293.49 & Hercules & 4620	 \\
NGC 7518 & 	 $ 0.07 $ & 	 $ 9.18 \pm 0.11 $ & 	 35.61 & Pegasus & 2590	 \\
NGC 7591 & 	 $ -0.49 $ & 	 $ 9.12 \pm 0.10 $ & 	 211.21 & Pegasus & 2480	 \\
NGC 7625 & 	 $ -0.42 $ & 	 $ 9.17 \pm 0.05 $ & 	 285.57 & UGCl 486 & 1880	 \\
NGC 7678 & 	 $ -0.05 $ & 	 $ 9.09 \pm 0.04 $ & 	 198.3 & Abell 2657 & 13100	 \\
\hline & & & & & \\
\multicolumn{6}{| c |}{\emph{Field Galaxies}} \\
\hline & & & & & \\
NGC 0095 & 	 $ -0.34 $ & 	 $ 8.96 \pm 0.07 $ & 	 203.78 & & 	 \\
NGC 0157 & 	 $ -0.34 $ & 	 $ 9.18 \pm 0.04 $ & 	 154.42 & & \\
NGC 0278 & 	 $ -0.05 $ & 	 $ 9.16 \pm 0.03 $ & 	 256.28 & &	 \\
NGC 0922 & 	 $ -0.42 $ & 	 $ 8.75 \pm 0.08 $ & 	 178.59 & &	 \\
NGC 1421 & 	 $ -0.26 $ & 	 $ 8.97 \pm 0.07 $ & 	 161.59 & &	 \\
NGC 2139 & 	 $ -0.54 $ & 	 $ 8.75 \pm 0.08 $ & 	 135.61 & &	 \\
NGC 2782 & 	 $ -0.05 $ & 	 $ 8.87 \pm 0.06 $ & 	 116.73 & &	 \\
NGC 2903 & 	 $ 0.08 $ & 	 $ 9.22 \pm 0.07 $ & 	 186.95 & &	 \\
NGC 3049 & 	 $ -0.05 $ & 	 $ 9.13 \pm 0.09 $ & 	 102.61 & &	 \\
NGC 3198 & 	 $ -0.14 $ & 	 $ 8.88 \pm 0.14 $ & 	 142.51 & &	 \\
NGC 3274 & 	 $ -0.60 $ & 	 $ 8.63 \pm 0.09 $ & 	 79.62 & &	 \\
NGC 3344 & 	 $ -0.07 $ & 	 $ 9.01 \pm 0.08 $ & 	 222.87 & &	 \\
NGC 3521 & 	 $ -0.18 $ & 	 $ 9.10 \pm 0.06 $ & 	 244.92 & &	 \\
NGC 3600 & 	 $ -0.28 $ & 	 $ 8.72 \pm 0.12 $ & 	 86.9 &	& \\
NGC 4384 & 	 $ -0.10 $ & 	 $ 9.03 \pm 0.05 $ & 	 102.29 & &	 \\
NGC 4455 & 	 $ -0.14 $ & 	 $ 8.71 \pm 0.10 $ & 	 56.98 & &	 \\
NGC 4605 & 	 $ 0.34 $ & 	 $ 8.81 \pm 0.07 $ & 	 60.87 & &	 \\
NGC 4670 & 	 $ -0.04 $ & 	 $ 8.72 \pm 0.08 $ & 	 140.4 & &	 \\
NGC 5104 & 	 $ -0.43 $ & 	 $ 8.81 \pm 0.19 $ & 	 203.18 & &	 \\
NGC 6207 & 	 $ -0.11 $ & 	 $ 8.97 \pm 0.05 $ & 	 114.83 & &	 \\
NGC 7137 & 	 $ -0.04 $ & 	 $ 9.12 \pm 0.04 $ & 	 104.46 & &	 \\
NGC 7620 & 	 $ -0.50 $ & 	 $ 9.07 \pm 0.05 $ & 	 423.69 & &	 \\
NGC 7624 & 	 $ -0.16 $ & 	 $ 9.21 \pm 0.07 $ & 	 173.88 & &	 \\
NGC 7640 & 	 $ -0.08 $ & 	 $ 8.84 \pm 0.09 $ & 	 107.92 & &	 \\
NGC 7742 & 	 $ 0.04 $ & 	 $ 9.08 \pm 0.05 $ & 	 112.06 & &	 \\

\end{supertabular}

\end{center}


\clearpage

\begin{thebibliography}{}
%
\bibitem[Baiesi-Pillastrini et al.(1984)]{bp84} Baiesi-Pillastrini, G.~C., Palumbo, G.~G.~C., \& Vettolani, G.\ 1984, \aaps, 56, 363
%
\bibitem[Cooper et al.(2008)]{cooper08} Cooper, M.~C., Tremonti, C.~A., Newman, J.~A. \& Zabludoff, A.~I. \ 2008, \mnras, 390, 245
%
\bibitem[Dav\'{e}(2008)]{dave08} Dav\'{e}, R. \ 2008, \mnras, 385, 147
%
\bibitem[Dav\'{e} et al.(2011a)]{dave11a} Dav\'{e}, R., Oppenheimer, B.~D., \& Finlator, K. \ 2011, \mnras, 415, 11
%
\bibitem[Dav\'{e} et al.(2011b)]{dave11b} Dav\'{e}, R., Finlator, K., \& Oppenheimer, B.~D. \ 2011, \mnras, 416, 1354
%
\bibitem[Dav\'{e} et al.(2012)]{dave12} Dav\'{e}, R., Finlator, K. \& Oppenheimer, B.~D. \ 2012, \mnras, 421, 98
%
\bibitem[Dav{\'e} et al.(2013)]{dave13} Dav{\'e}, R., Katz, N., Oppenheimer, B.~D., Kollmeier, J.~A., \& Weinberg, D.~H.\ 2013, arXiv:1302.3631
%
\bibitem[de Vaucouleurs et al.(1991)]{gdv91} de Vaucouleurs, G., de Vaucouleurs, A., Corwin, H.~G., Jr., et al.\ 1991, Third Reference Catalogue of Bright Galaxies.~Volume I: Explanations and references.~ Volume II: Data for galaxies between 0$^{h}$ and 12$^{h}$.~ Volume III: Data for galaxies between 12$^{h}$ and 24$^{h}$., by de Vaucouleurs, G.; de Vaucouleurs, A.; Corwin, H.~G., Jr.; Buta, R.~J.; Paturel, G.; Fouqu{\'e}, P..~Springer, New York, NY (USA), 1991, 2091 p., ISBN 0-387-97552-7, Price 
US\$ 198.00.~ISBN 3-540-97552-7, Price DM 448.00.~ISBN 0-387-97549-7 (Vol.~I), ISBN 0-387-97550-0 (Vol.~II), ISBN 0-387-97551-9 (Vol.~III).
%
\bibitem[Duffy et al.(2012)]{duffy12} Duffy, A.~R., Kay, S.~T., Battye, R.~A. et al. \ 2012, \mnras, 420, 2799
%
\bibitem[Ellison et al.(2008)]{ellison08} Ellison, S.~L., Patton, D.~R., Simard, L. \& McConnachie, A.~W. \ 2008, \aj, 135, 1877
%
\bibitem[Ellison et al.(2009)]{ellison09} Ellison, S.~L., Simard, L., Cowan, N.~B. et al. \ 2009, \mnras, 396, 1257
%
\bibitem[Finlator \& Dav\'{e}(2008)]{finlator08} Finlator, K. \& Dav\'{e}, R. \ 2008, \mnras, 385, 2181
%
\bibitem[Howell et al.(2010)]{howell10} Howell, J.~H., Armus, L., Mazzarella, J.~M., et al.\ 2010, \apj, 715, 572
%
\bibitem[Giovanelli \& Haynes(1985)]{giovanelli85} Giovanelli, R., \& Haynes, M.~P.\ 1985, \apj, 292, 404 
%
\bibitem[Gunn \& Gott(1972)]{gunn72} Gunn, J.~E., \& Gott, J.~R. \ 1972, \apj, 176, 1
%
\bibitem[Isobe et al.(1990)]{isobe90} Isobe, T., Feigelson, E.~D., Akritas, M.~G., \& Babu, G.~J. \ 1990, \apj, 364, 104
%
\bibitem[Kelly(2007)]{kelly07} Kelly, B.~C. \ 2007, \apj, 665, 1489
%
\bibitem[Kennicutt(1998)]{kennicutt98} Kennicutt, R.~C., Jr. \ 1998, \apj, 498, 541
%
\bibitem[Kewley et al.(2006)]{kewley06} Kewley, L.~J., Geller, M.~J., \& Barton, E.~J. \ 2006, \aj, 131, 2004
%
\bibitem[Levy et al.(2007)]{levy07} Levy, L., Rose, J.~A., van Gorkom, J.~H. \& Chaboyer, B. \ 2007, \aj, 133, 1104
%
\bibitem[Moustakas \& Kennicutt(2006a)]{moustakas06} Moustakas, J. \& Kennicutt, R.~C., Jr. \ 2006, \apjs, 164, 81
%
\bibitem[Moustakas \& Kennicutt(2006b)]{mk06} Moustakas, J. \&
  Kennicutt, R.~C., Jr. \ 2006, \apj, 651, 155
%
\bibitem[Paturel et al.(2003)]{paturel03} Paturel, G., Petit, C., Prugniel, Ph. et al. \ 2003, \aap, 412, 45
%
\bibitem[Petropoulou et al.(2012)]{petropoulou12} Petropoulou, V., V\'{i}lchez, J., \& Iglesias-P\'{a}ramo, J. \ 2012, \apj, 749, 133
%
\bibitem[Popping et al.(2009)]{popping09} Popping, A., Dav\'{e}, R., Braun, R. \& Oppenheimer, B.~D. \ 2009, \aap, 504, 15
%
\bibitem[Robertson et al.(2012)]{robertson12} Robertson, P., Shields,
  G.~A., \& Blanc, G.~A. \ 2012, \apj, 748, 48
%
\bibitem[Rose et al.(2010)]{rose10} Rose, J.~A., Robertson, P., Miner, J. \& Levy, L. \ 2010, \aj, 139, 765
%
\bibitem[Skillman et al.(1996)]{skillman96} Skillman, E.~D., Kennicutt, R.~C., Jr., Shields, G.~A., \& Zaritsky, D. \ 1996, \apj, 462, 147
%
\bibitem[Solanes et al.(1996)]{sgh96} Solanes, J.~M., Giovanelli, R., \& Haynes, M.~P. \ 1996, \apj, 461, 609
%
\bibitem[Solanes et al.(2001)]{solanes01} Solanes, J.~M., Manrique, A., Garc{\'{\i}}a-G{\'o}mez, C., et al.\ 2001, \apj, 548, 97
%
\bibitem[Tremonti et al.(2004)]{tremonti04} Tremonti, C.~A. \etal\ 2004, \apj, 613, 898
%
\bibitem[Zaritsky et al.(1994)]{zkh94} Zaritsky, D., Kennicutt, R.~C., Jr., \& Huchra, J.~P. \ 1994, \apj, 420, 87
%
\bibitem[Zhang et al.(2009)]{zhang09} Zhang, W., Li, C., Kauffmann, G. et al. \ 2009, \mnras, 397, 1243
%
\end{thebibliography}
\end{document}